\documentclass{elsart}

\usepackage{graphicx}
\usepackage{amssymb}

\begin{document}

\begin{frontmatter}

\title{Electronic structure of the Au/benzene-1,4-dithiol/Au transport
interface: Effects of chemical bonding}

\author{U.~Schwingenschl\"ogl\corauthref{cor1}},
\ead{Udo.Schwingenschloegl@physik.uni-augsburg.de}
\author{C.~Schuster}
\corauth[cor1]{Corresponding author. Fax: 49-821-598-3262}
\address{Institut f\"ur Physik, Universit\"at Augsburg, 86135 Augsburg,
Germany}

\begin{abstract}
We present results of electronic structure calculations for well-relaxed
Au/benzene-1,4-dithiol/Au molecular contacts, based on density functional
theory and the generalized gradient approximation. Electronic states
in the vicinity of the Fermi energy are mainly of Au $5d$
and S $3p$ symmetry, whereas contributions of C $2p$ states are very small.
Hybridization between C $2p$ orbitals within the benzene substructure is
strongly suppressed due to S-C bonding. In agreement with
experimental findings, this corresponds to a significantly reduced
conductance of the molecular contact.

\end{abstract}

\end{frontmatter}

\section{Introduction}

Molecular devices nowadays are becoming more and more important because
of a rapidly increasing number of potential applications, spanning a wide
range from new computer architectures to multifunctional chemical sensors
and medical diagnostic tools. Exploring the use of specific molecules as
active components in electronic circuits has been a field of special
interest in recent years. Electrical properties of molecular devices are
usually addressed by measuring the transport through a molecule sandwiched
between metallic electrodes. The resulting transport characteristics, of
course, reflects both the metal-molecule contact and the molecule itself.
Because transport in general is restricted to a small number of valence
orbitals, the features of a molecular device strongly depend on the local
electronic structure at the different molecular sites. Details of the
chemical bonding, such as hybridization or charge transfer, hence play
a key role for understanding the device, and structural relaxation due to
the metal-molecule contact has to be taken into consideration.

Experimentally, Au/benzene-1,4-dithiol/Au transport interfaces are
prepared via the mechanically controllable break junction technique
\cite{reed97}. In a first step a thin gold wire, covered with a
self-assembled benzene-1,4-dithiol monolayer, is streched until breakage.
The opposing gold contacts then are slowly moved together until the onset
of conductance is reached. From the theoretical point of view, the
prototypical Au/benzene-1,4-dithiol/Au interface likewise has attracted
much attention. Band structure methods have succeeded in reproducing
the shape of the experimental current-voltage characteristic of
the contact when an external bias is applied. However, the absolute
value of the conductance is overestimated by one or two orders of
magnitude \cite{ventra00,xue03,garcia06}. The current magnitude appears
to be an extremely sensitive function of both the contact geometry and
chemistry. But even if atomic details of the contact are taken into
account, transport calculations fail to reproduce the experimental
fact of a conductance gap at zero bias.

The benzene-1,4-dithiol molecule has $\pi$ bonding and $\pi^*$ antibonding
orbitals formed by the carbon and sulfur $p$ orbitals perpendicular to the
ring plane, as well as $\sigma$ bonding and $\sigma^*$ antibonding orbitals
due to in-plane orbital overlap. While both the $\sigma$ and $\pi$ orbitals
are fully occupied, the antibonding orbitals stay empty. Because they
mediate direct orbital overlap, $\sigma$ states are subject to stronger
bonding-antibonding splitting than $\pi$ states, and therefore appear at
lower energies. For a small bias, the conductance of the molecular contact
shows Ohmic behaviour because of a smooth density of states near the Fermi
energy. When the bias is increased, two conduction peaks are observed, which
can be traced back to resonant tunneling through either $\pi^*$ or
(at higher bias) $\pi$ orbitals \cite{ventra00}.

The erroneous prediction of metallic transport characteristics for
experimentally insulating molecules has been associated with the
continuous exchange-correlation approximation applied in standard
density-functional schemes \cite{toher05}. For the
Au/benzene-1,4-dithiol/Au contact, it has been demonstrated that an
atomic self-interaction correction is useful to compensate this shortcoming
in an approximate way, opening a conduction gap in the current-voltage
characteristic. However, in this letter we show that the modification
of the chemical bonding in the benzene ring, resulting from the specific
contact geometry, has serious effects on the local electronic structure
and thus likewise explains the insulating state. We thus conclude that
taking into account the very details of the crystal and electronic
structure is mandatory in order to obtain adequate input for transport
calculations.

\section{Calculational Details}

The electronic structure calculations presented in the following are
based on density functional theory and the generalized gradient
approximation. To be specific, we use the WIEN2k program package,
a state-of-the-art full-potential code applying a mixed lapw and apw+lo
basis \cite{wien2k}. In particular, the
WIEN2k program allows for structural optimization of the interface
geometry. Thus it is suitable to study
effects of covalent bonding and hybridization in detail. A well-relaxed
interface geometry is of great importance for reliable band structure
results, as even small alterations of the orbital overlap, due to
structural relaxation effects, may cause significant modifications of the
electronic states \cite{us03,eyert04,schmitt05,eyert05,us07}.

In our calculations, the charge density is represented via 88491 plane
waves, the {\bf k}-mesh for the Brillouin zone integrations comprises
72 {\bf k}-points, and the Perdew-Burke-Ernzernhof parametrization of the
exchange-correlation potential is applied. While Au $5p$, S $3s$, and C $2s$
orbitals are treated as semi-core states, the valence states comprise
Au $5d$, $6s$, $6p$, S $3p$, $4s$, C $2p$, $3s$ as well as H $1s$ orbitals.
Corresponding atomic sphere radii amount to $2.6\,a_B$ for the Au,
$1.8\,a_B$ for the S, $1.3\,a_B$ for the C, and $0.7\,a_B$ for the
H spheres.

\begin{figure}[h]
\centering 
\includegraphics*[width=0.3\textwidth]{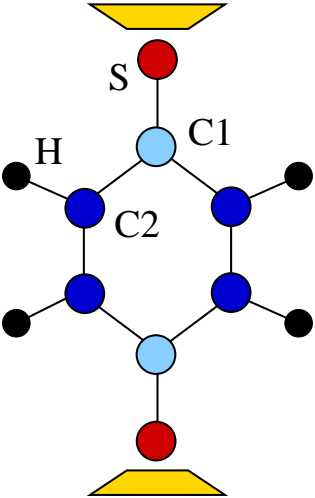}
\caption{Schematic structure of the Au/benzene-1,4-dithiol/Au molecular
contact.}
\label{fig1}
\end{figure}
\vspace{0.5cm}
\begin{table}[h]
\begin{tabular}{l|c|c|c|c|c}
&Au-S&S-C1&C1-C2&C2-C2&C2-H\\\hline
bond length&2.55\,\AA&1.76\,\AA&1.42\,\AA&1.38\,\AA&1.10\,\AA
\end{tabular}
\vspace{0.5cm}
\caption{Bond lengths in the relaxed Au/benzene-1,4-dithiol/Au molecular
contact.}
\label{tab1}
\end{table}

Our calculation relies on the contact geometry shown in figure \ref{fig1}.
The benzene-1,4-dithiol molecule is sandwiched between two gold contacts
in fcc [001] orientation. While the carbon site C2 forms the center of the
molecule, the site C1 contacts the sulfur atom on top of a gold pyramide.
Each sulfur atom thus has four nearest neighbour gold sites.
Moreover, the fourth gold layer off the contact forms the basal plane of
our unit cell, which allows us to use periodic boundary conditions.
In total, each electrode consequently comprises 23 gold atoms in four layers.
A convenient starting point for the contact geometry is given by the
bulk gold bond lengths and angles, the bond lengths of the bare molecule,
and a typical Au-S distance. Structural relaxation is carried out by means
of minimization of interatomic forces as well as the total energy. The
resulting bond lengths for the optimized contact geometry, as summarized
in table \ref{tab1}, largely resemble the numbers reported by Basch and
Ratner \cite{basch03}. Our subsequent discussion of the
Au/benzene-1,4-dithiol/Au molecular contact is based on electronic
structure data for the fully relaxed geometry.

\section{Results and Discussion}

\begin{figure}
\centering 
\includegraphics*[width=0.5\textwidth]{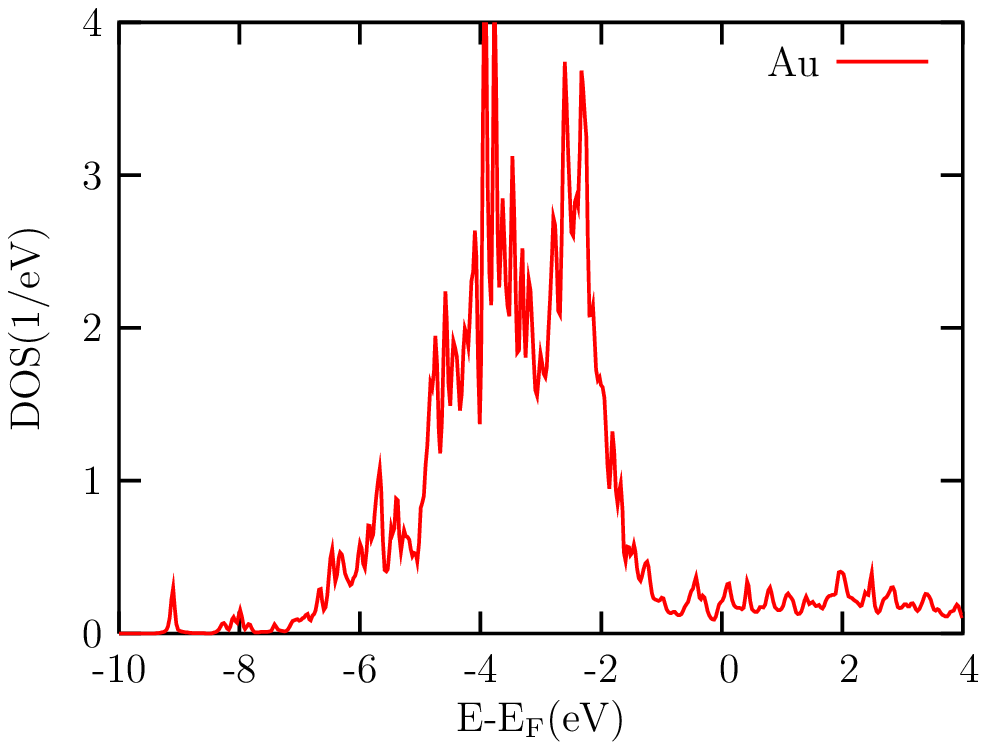}\\[0.2cm]
\includegraphics*[width=0.5\textwidth]{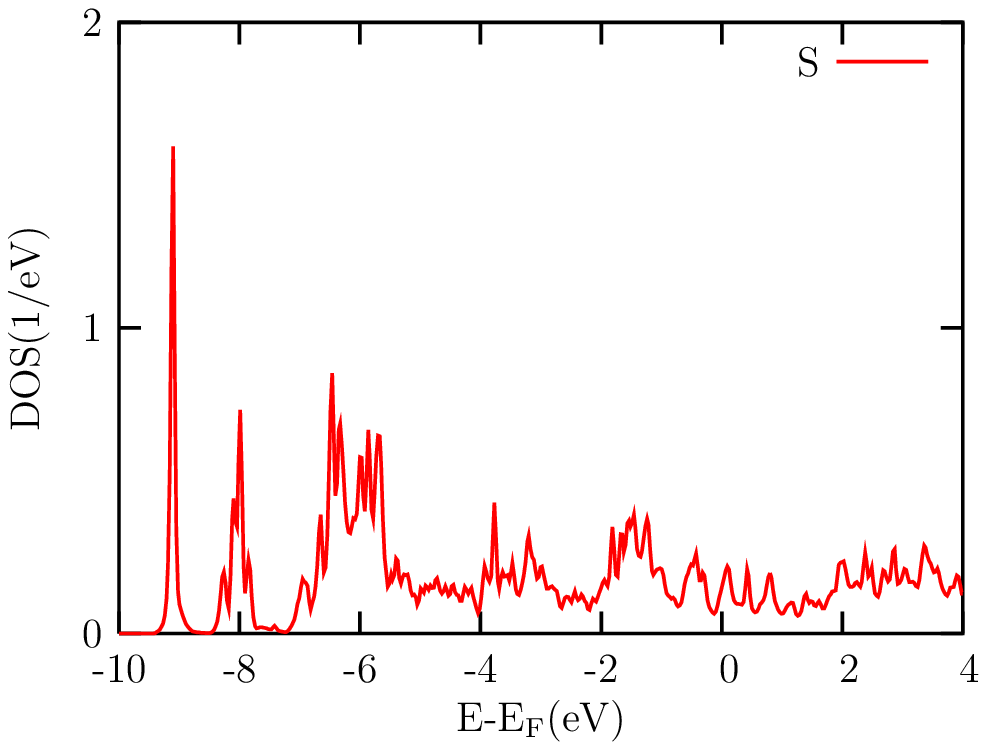}
\caption{Au/benzene-1,4-dithiol/Au molecular contact:
Partial Au and S densities of states. The DOS is
normalized with respect to the number of contributing sites.}
\label{fig2}
\end{figure}

\begin{figure}
\centering 
\includegraphics*[width=0.5\textwidth]{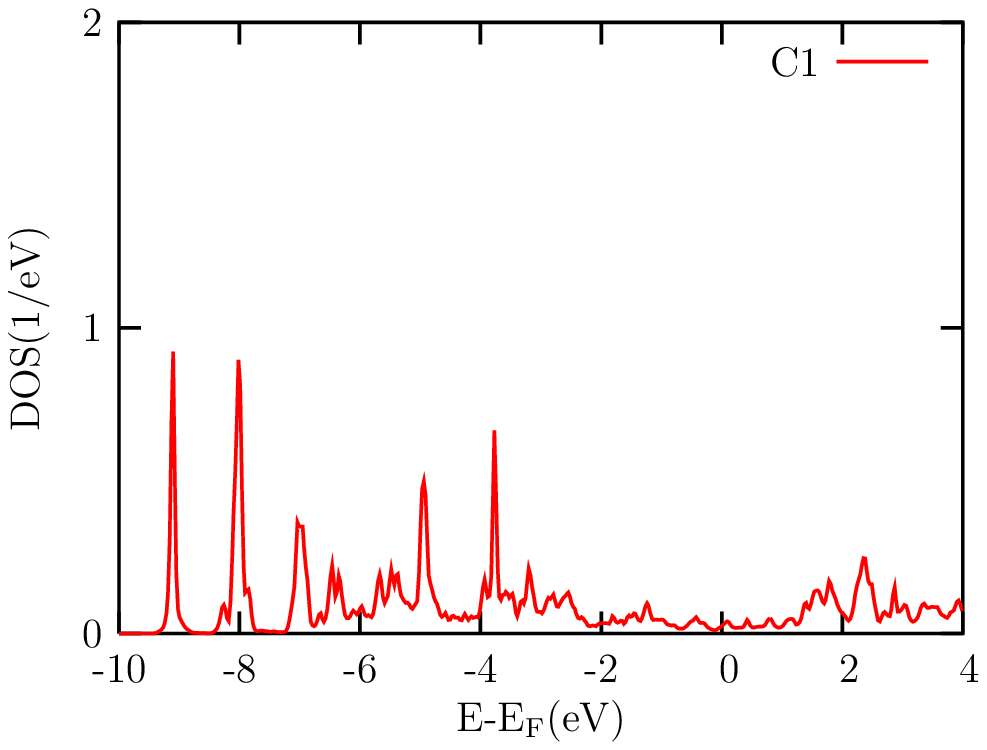}\\[0.2cm]
\includegraphics*[width=0.5\textwidth]{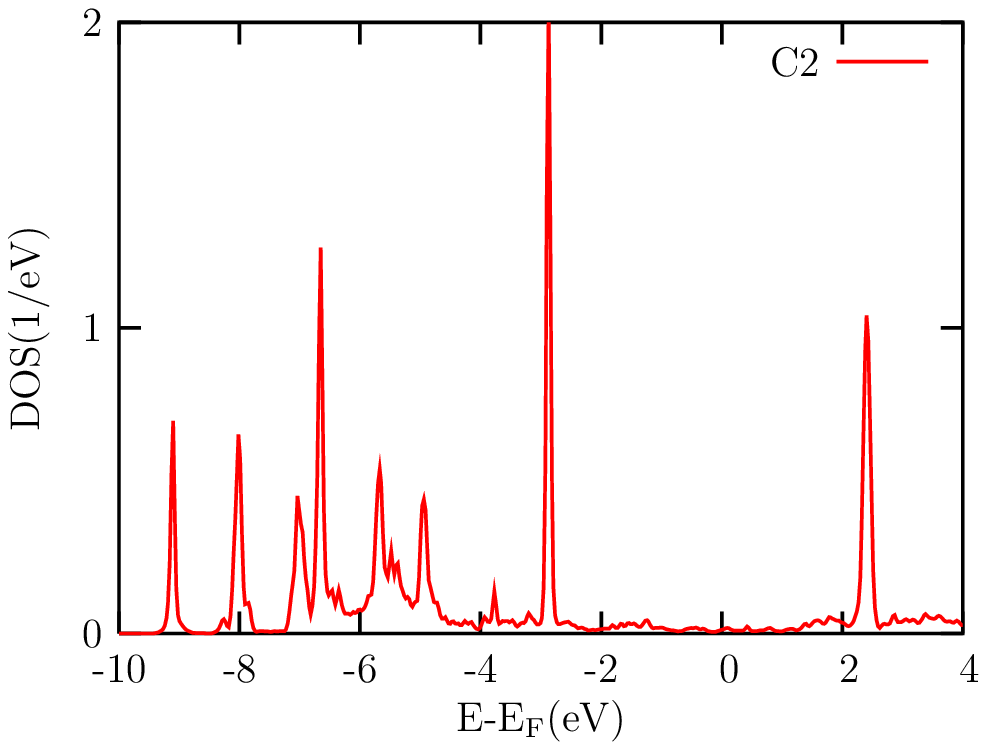}
\caption{Au/benzene-1,4-dithiol/Au molecular contact:
Partial C densities of states for carbon sites 1 and 2 (see figure
\ref{fig1}). The DOS is normalized with respect to the number of
contributing sites.}
\label{fig3}
\end{figure}

Results of our LAPW calculation for the Au/benzene-1,4-dithiol/Au
interface are shown in figures \ref{fig2} and \ref{fig3}. We present
partial densities of states (DOS) for the contact Au sites (forming the
top of the gold electrodes), the S sites, and the two carbon sites C1 and C2;
compare the structure sketch in figure \ref{fig1}. For a
quantitative analysis, the density of states is normalized with respect to the number
of contributing sites in each case. In the energy range shown, almost all
electronic states trace back to Au $5d$, S $3p$, and C $2p$ orbitals,
respectively. The partial DOS of the contact Au sites is dominated by a
broad structure of Au $5d$ states, which is centered at roughly $-3.5$\,eV
and extends far above the Fermi energy. The gross features of the DOS
shape resemble the findings for fcc gold, reflecting metallic conductivity
for these sites. Moreover, hardly any deviation from fcc gold is found
for the electronic structure of Au sites in the third gold layer off
the contact, which justifies the setup of our electrodes.

The S $3p$ DOS is rather unstructured in a wide energy window reaching
from about $-5$\,eV to $4$\,eV. Especially, at the Fermi level a distinct
DOS is present, with a magnitude similar to the Au $5d$ DOS. This
observation traces back to hybridization between gold and sulfur orbitals,
as expected for Au-S bonding.

In the next step we turn to the C1 site,
which is affected by S-C1 and C1-C2 bonding. As compared to the previous
findings for gold and sulfur, we have a significantly reduced (but still
finite) DOS in the vicinity of the Fermi energy. Moreover, strong
hybridization between C1 $2p$ and S $3p$ orbitals is evident, as both
densities of states have sharp peaks near $-9$\,eV and $-8$\,eV, and
common structures around both $-6$\,eV and $-4$\,eV. Hence, remarkable
S-C1 bonding is a typical feature of the benzene-1,4-dithiol molecule.

Sharp DOS peaks at $-9$\,eV and $-8$\,eV and a distinct structure reaching
from about $-7$\,eV to $-5$\,eV likewise are present for the C2
site. These states belong to $\sigma$ bonding orbitals of $p$ symmetry,
which is also true for the corresponding S and C1 states. Of course,
$\sigma$ bonding orbitals mediate the main part of the orbital overlap
within the benzene ring and with adjacent sulfur atoms. They therefore
are subject to strong hybridization. States at energies above $-4$\,eV are
expected to belong to $\pi$ bonding and $\pi^*$ antibonding orbitals. In
fact, the C2 $2p$ DOS has characteristic peaks near $-3$\,eV and $2.5$\,eV,
which, however, are not found in the C1 $2p$ DOS. Hybridization between
$\pi$ bonding orbitals at sites C1 and C2 hence is almost negligible. As
a consequence, $\pi$-type coupling within the benzene ring is seriously
reduced in case of the benzene-1,4-dithiol molecule, which we attribute
to strong S-C1 bonding. Due to the latter, C1 $\pi$ and $\pi^*$ states
are subject to energetical shifts, yielding a rather unstructured DOS at
energies above $-4$\,eV. Since $\pi$ bonding states are responsible for
the electrical conductance of the benzene ring, reduced C1-C2 coupling
leaves the whole ring insulating. In fact, the C2 $2p$ DOS reveals hardly
any states in the vicinity of the Fermi energy.

In conlusion, we have studied the electronic structure of the
Au/benzene-1,4-dithiol/Au transport interface by means of band structure
calculations within density functional theory. Taking into account the
specific interface geometry, we have addressed the chemical bonding in
the molecular contact. While coupling via $\sigma$ bonding S $3p$ and
C $2p$ orbitals is strong, hybridization between $\pi$ bonding orbitals
at neighbouring carbon sites is almost completely suppressed due to S-C
bonding. As a consequence, the local DOS at carbon sites without sulfur
bonding almost vanishes at the Fermi energy, which explains the
experimental observation of an insulating state. A transport calculation
neglecting the details of the chemical bonding in the benzene-1,4-dithiol
molecule is expected to overestimate the conductance (at least at small
bias), although the shape of the current-voltage characteristic may be
predicted correctly \cite{ventra00,xue03,garcia06}.

\section*{Acknowledgements}
We thank U.\ Eckern and P.\ Schwab for fruitful discussions, and
acknowledge financial support by the Deutsche Forschungsgemeinschaft
within SFB 484.

\end{document}